\documentclass[a4paper,10pt,twocolumn]{article}
\usepackage[english]{babel}
\usepackage{graphicx}
\usepackage{fixltx2e}
\usepackage[hypertex]{hyperref}
\usepackage{amsmath}
\usepackage{longtable}
\usepackage{mathrsfs}
\usepackage{amsmath}
\usepackage{soul}
\usepackage{color}
\usepackage{textcomp}
\usepackage{abstract}
\graphicspath{{figures/}}

\usepackage[small]{caption2}

\begin{document}

\title{Diffraction microtomography with sample rotation: influence of a missing apple core in the recorded frequency space}
\author{Stanislas Vertu\textdagger, Jean-Jacques Delaunay$^{*}$\textdagger, Olivier Haeberle\textdaggerdbl\\
\small \textdagger\,Department of Engineering Synthesis, School of Engineering, The University\\[-0.8ex]
\small of Tokyo, 7-3-1 Hongo Bunkyo-ku, Tokyo 113-8656, Japan\\
\small \textdaggerdbl\,Laboratory MIPS \textendash\, University of Haute Alsace, IUT Mulhouse, 61 rue\\[-0.8ex]
\small Albert Camus, 68093 Mulhouse Cedex, France\\
\small $^{*}$ \texttt{E-mail: jean@mech.t.u-tokyo.ac.jp}}

\date{\small{(dated: May 5, 2008)}}

\makeatletter
\newenvironment{tablehere}
  {\def\@captype{table}}
  {}

\newenvironment{figurehere}
  {\def\@captype{figure}}
  {}

\renewenvironment{thebibliography}[1]
     {\section*{\refname}%
      \@mkboth{\MakeUppercase\refname}{\MakeUppercase\refname}%
      \list{\@biblabel{\@arabic\c@enumiv}}%
           {\settowidth\labelwidth{\@biblabel{#1}}%
            \leftmargin\labelwidth
            \advance\leftmargin\labelsep
            \@openbib@code
            \usecounter{enumiv}%
            \let\p@enumiv\@empty
            \renewcommand\theenumiv{\@arabic\c@enumiv}}%
      \setlength{\itemsep}{0pt}
      \sloppy
      \clubpenalty4000
      \@clubpenalty \clubpenalty
      \widowpenalty4000%
      \sfcode`\.\@m}
     {\def\@noitemerr
       {\@latex@warning{Empty `thebibliography' environment}}%
      \endlist}

\makeatother

\renewcommand{\thesection}{\arabic{section}.}

\twocolumn[
\maketitle 
\begin{onecolabstract}
Diffraction microtomography in coherent light is foreseen as a promising technique to image transparent living samples in three dimensions without staining. Contrary to conventional microscopy with incoherent light, which gives morphological information only, diffraction microtomography makes it possible to obtain the complex optical refractive index of the observed sample by mapping a three-dimensional support in the spatial frequency domain. The technique can be implemented in two configurations, namely, by varying the sample illumination with a fixed sample or by rotating the sample using a fixed illumination. In the literature, only the former method was described in detail. In this report, we precisely derive the three-dimensional frequency support that can be mapped by the sample rotation configuration. We found that, within the first-order Born approximation, the volume of the frequency domain that can be mapped exhibits a missing part, the shape of which resembles that of an apple core. The projection of the diffracted waves in the frequency space onto the set of sphere caps covered by the sample rotation does not allow for a complete mapping of the frequency along the axis of rotation due to the finite radius of the sphere caps. We present simulations of the effects of this missing information on the reconstruction of ideal objects.
\end{onecolabstract}
\bigskip
]

\section{Introduction}

In transmission microscopy, the observation of non-fluorescent transparent or quasi-transparent samples poses a real challenge because of the inherent low contrast in refractive index associated with this type of samples. This low contrast makes it difficult to identify intracellular structures. Furthermore, conventional transmission microscopes suffer from a poor resolution, especially along the optical axis because of incomplete mapping of the observed object frequencies. The missing information, referred to as the \textquotedblleft missing cone\textquotedblright, induces anisotropy in the spatial resolution which is responsible for object deformation. To address these two issues, sample rotation and subsequent observation of the sample under different view angles can be used to record a more complete set of the object frequencies and reconstruct the three-dimensional object from this extended frequency map. For non-diffractive samples, this method is well-known in the medical field under the name of Computed Tomography (CT). For weakly diffractive samples, the three-dimensional distribution of the complex refractive index can be reconstructed from the knowledge of the scattered fields sampled under various view angles according to the diffraction tomography theorem \cite{Wolf1969,Dandliker1970}. Thus the observation of cell internal structures as well as cell arrangements in the three dimensions is in principle possible.

Three-dimensional observation of weakly diffractive samples have been so far realized by varying the sample illumination with a fixed sample \cite{Lauer2002,Kawata1987,Haeberle2007,Simon2007} or by rotating the sample using a fixed illumination \cite{Depeursinge2006a}. In the sample illumination variation method, tomography is realized by varying the sample illumination angle in a range limited by the illumination system (usually a condenser) numerical aperture, drastically increasing the object spatial frequencies captured by this system in comparison to a conventional holographic transmission microscope, which uses only one direction of illumination. The three-dimensional refractive index distribution of the sample is reconstructed by applying the diffraction tomography theorem. Because of the limited range of illumination angles, the object frequencies are not captured isotropically, a \textquotedblleft missing cone\textquotedblright\, remains and the reconstruction exhibits deformation along the optical axis.

In the sample rotation method, the sample is inserted in a microcapillary, which is used to precisely rotate the sample. This configuration allows for a complete rotation of the sample, but unavoidable displacements of the microcapillary may induce errors in the sample position, and therefore in the image reconstruction. The reconstruction of the variation in refractive index of transparent samples is usually obtained by backprojecting the measured sample phase using the Radon transform. By applying backprojection, the obtained spatial resolution is isotropic and no information is missing along the rotation axis. However this reconstruction by backprojection assumes negligible diffraction. A more complete treatment that includes the effects of weak scattering from the sample can be obtained using the diffraction tomography theorem. In this study, we derive the object frequency support mapped by diffraction tomography with sample rotation of a weakly scattering sample.

We found that diffraction tomography with sample rotation offered a nearly isotropic spatial resolution. Missing object frequencies were found to be restricted along the axis of rotation with a shape of the missing frequency support that resembles that of an apple core, therefore the chosen name of \textquotedblleft missing apple core\textquotedblright.

The non-diffractive reconstruction technique used in the backprojection of the phase is first reviewed in Section 2, then the tomography diffraction reconstruction technique is used in Section 3 to derive the support of the recorded object frequencies. Finally the effects of the missing object frequencies on the reconstruction of ideal objects are examined in Section 4.

\section{Non-diffractive optical coherent microtomography}

\begin{figure}[!t]
\centering
\includegraphics[width=0.48\textwidth]{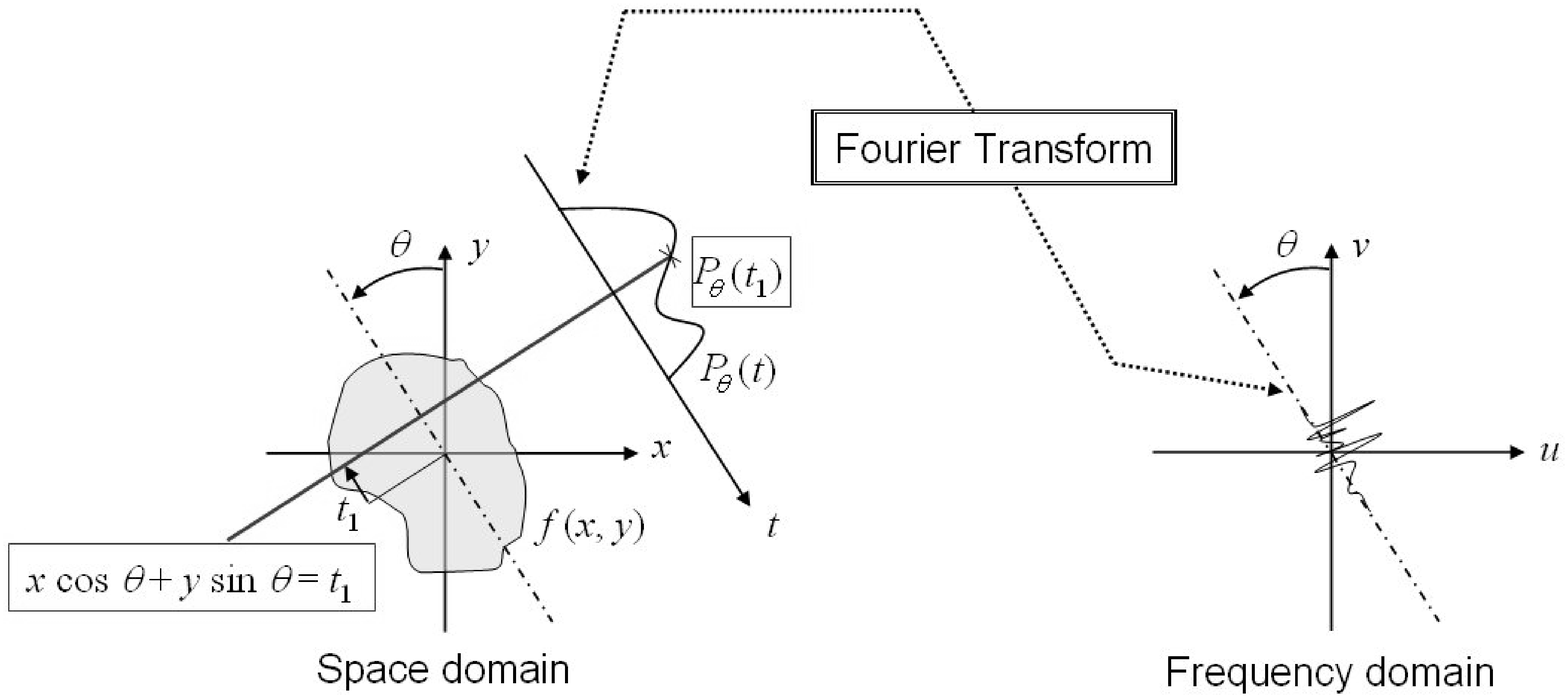}
\caption{Schematic of the Fourier slice theorem showing the relation between the Fourier transform of the projected data of the object $f(x,y)$ and the Fourier transform of the same object. The Fourier transform values of the projected data are distributed along a straight radial line.}
\label{fig:1}
\end{figure}
The backprojection reconstruction technique widely used in CT provides a means to reconstruct non-diffractive three-dimensional objects from the intensity measurement under different view angles. For non-diffractive transparent objects, the measurement of the object phase under different view angles and the subsequent phase backprojection make possible the reconstruction of the distribution of the refractive index difference in three dimensions.

For weakly diffractive transparent objects, the assumption of negligible diffraction may be a useful approximation for objects exhibiting a slow variation in refractive index, as shown by results in \cite{Depeursinge2006a,Vertu2008}. The condition of a small variation in refractive index is satisfied by optically matching the sample with an index matching material. It should be noted however that the backprojection technique is strictly valid for non-diffractive objects only and that neglecting diffraction generates artifacts in observed images as illustrated in a recent work \cite{Gorski2007} where circular apertures in a hollow fiber (two-dimensional photonic crystal) were reconstructed with triangular shapes when using the backprojection technique.

Theoretically, for one angle of illumination and one ray of light, the intensity measured is related to the object by a line integral as shown in Figure \ref{fig:1}. Let's denote the object by a two-dimensional function $f(x,y)$ and define the line integrals by the $(\theta,t)$ parameters. Using a delta function, the Radon transform $P_\theta(t)$ of the function $f(x,y)$ can be written as:
\begin{eqnarray}
P_\theta (t)= \!\iint \limits_{\:-\infty\:-\infty}^{\:+\infty\:+\infty} \!\!\!\! f(x,y) \delta(x\cos \theta +y\sin \theta -t) \, \mathrm{d}x \, \mathrm{d}y.
\label{eq1}
\end{eqnarray}

The sample is illuminated by a parallel plane wave, so the projection is formed by combining a set of parallel ray integrals. The Fourier slice theorem relates the one-dimensional Fourier transform of a parallel projection $P_\theta(t)$ of an image $f(x,y)$ taken at an angle $\theta$ to a slice of the two-dimensional transform $F(u,v)$ of this image, subtending an angle $\theta$ with the $u$-axis. In other words, the Fourier transform of $P_\theta(t)$ gives the values of $F(u,v)$ along the dashed line represented in the frequency domain on the right side of Figure \ref{fig:1}.

Moreover a real microtomography optical setup has a limited Numerical Aperture ($NA$) imposed by the microscope objective or by the complete system. Consequently the setup acts as a three-dimensional rotational-symmetric low-pass filter. Therefore the measured data in the frequency domain are limited by a disk for one direction of illumination. By applying the Fourier slice theorem with a fine enough step for $\theta$, it is possible to isotropically map an extended support of object frequencies that completely fills a sphere.
\begin{figure}[!t]
\centering
\includegraphics[width=0.48\textwidth]{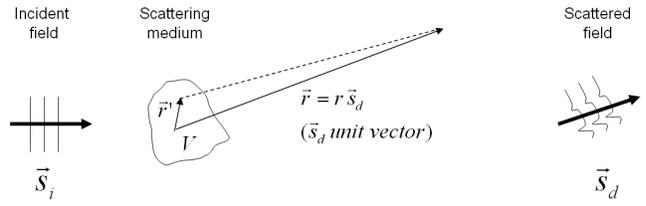}
\caption{Definition of the propagation vector of the scattered light from a medium of volume $V$.}
\label{fig:2}
\end{figure}

\section{Diffractive optical coherent microtomography}

Diffraction by weakly scatterers can be treated within the framework of the first-order Born approximation that gives an approximated solution to the inverse scattering problem. The situation is now depicted in Figure \ref{fig:2}, where $U^{(i)}$ is the incident field and $U^{(s)}$ is the scattered field. The field incident on the weakly scattering object is taken as a monochromatic plane wave of unit amplitude and wavelength $\lambda$, propagating in the direction specified by the wave vector $\vec{k_i}=k\: \vec{s_i}$ where $k = 2\pi / \lambda$ and $\vec{s_i}$ is a unit vector, so the expression for the incident field $U^{(i)}(r)=e^{i\vec{k_i}\cdotp\vec{r}}$.

The scattered field propagating in the direction $\vec{s_d}$ has a wave vector $\vec{k_d}=k\: \vec{s_d}$. Born and Wolf \cite{Wolf1969,BornWolf} have shown that within the first-order Born approximation and in the far zone approximation ($k\:r\rightarrow\infty$), the expression for the scattering field $f(\vec{s_d},\vec{s_i})$ is:
\begin{eqnarray}
f(\vec{s_d},\vec{s_i})= \iiint \limits_{V} F(\vec{r}\,') e^{-ik(\vec{s_d}-\vec{s_i})\cdotp\vec{r}\,'} \, \mathrm{d}^3\vec{r}\,' ,
\label{eq2}
\end{eqnarray}
where $F(\vec{r}\,')$ r is the scattering potential of the medium which varies with the square of the complex refractive index of the medium and $\vec{r}\,'$ is a vector describing the volume $V$ of the scatterer.

By introducing the vector $\vec{K}=k(\vec{s_d}-\vec{s_i})$ in Equation \ref{eq2}, the scattered field can be expressed as the Fourier transform of the scattering potential evaluated at $\vec{K}$:
\begin{eqnarray}
f(\vec{s}_d,\vec{s}_i) = \tilde{F}(\vec{K}) = \iiint \limits_V F(\vec{r}\,')  e^{-i\vec{K}\cdot\vec{r}\,'} \,\mathrm{d}^3\vec{r}\,'.
\label{eq3}
\end{eqnarray}
The vector $\vec{K}$ describes the different subsets for which the scattering potential frequency can be obtained from the scattered field.

\begin{figure}[!t]
\centering
\includegraphics[width=0.48\textwidth]{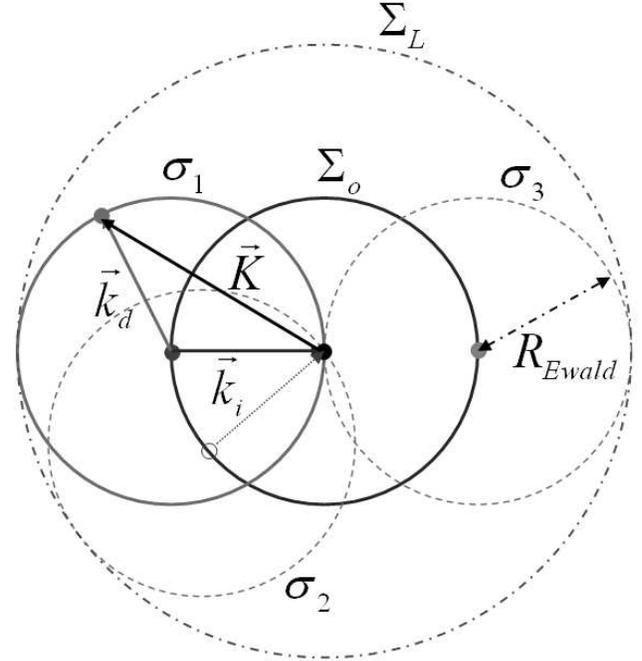}
\caption{Two-dimensional representation of the Ewald limiting sphere $\Sigma_L$ and the support of the Ewald sphere $\Sigma_0$, the radius of which is the wavenumber of the incident light. $\Sigma_0$ represents all possible directions of incidence and $\Sigma_L$ is the largest frequency domain covered by the scattered vectors when all possible incidence directions are mapped. A $\sigma$ Ewald's sphere represents all possible scattered directions for one illumination direction. Note that a transmission microscope collects at best half of the $\sigma$ spheres.}
\label{fig:3}
\end{figure}

The region of the frequency domain described by the $\vec{K}$-vectors forms a sphere called the Ewald sphere, which is shown in Figure \ref{fig:3}. Equation \ref{eq3} provides a very powerful relation between a quantity that can be determined by measurements $f(\vec{s}_d,\vec{s}_i)$ and a quantity that is related to the complex refractive index of the object $\tilde{F}(\vec{K})$. In other words, by illuminating a weakly scattering object by a plane wave having direction $\vec{s}_i$, the knowledge of the scattered field in direction $\vec{s}_d$ in the far field and the incident field allows to reconstruct a three-dimensional subset of the object refractive index distribution in the frequency domain. The measure of the scattered field in all possible illumination directions and all possible diffraction directions makes it possible to completely cover the Ewald limiting sphere shown in Figure \ref{fig:3} and therefore obtain an isotropic spatial resolution.

\begin{figure}[!t]
\centering
\includegraphics[width=0.48\textwidth]{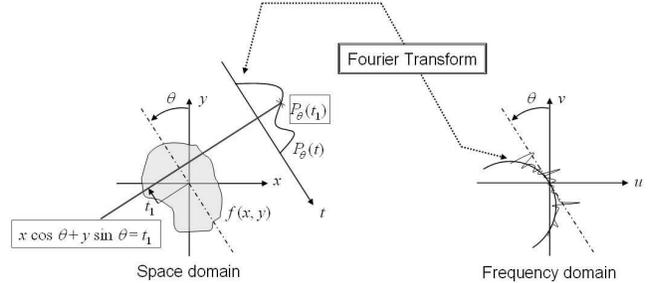}
\caption{Schematic of the Fourier diffraction theorem showing the relation between the Fourier transform of the projected data of the object $f(x,y)$ and the Fourier transform of the same object. In contrast to the non-diffracting case of Figure \ref{fig:1}, the Fourier transform values of the projected data are distributed along an arc of a circle. Note that the theorem is valid under the first-order Born approximation.}
\label{fig:4}
\end{figure}

Considering now a microtomography \textit{diffractive} setup in transmission, the relation between the measured scattered field and the Fourier transform of the projected data is shown in Figure \ref{fig:4}. In contrast to the backprojection case of Figure \ref{fig:1} where the Fourier transform values lay on a straight line, the Fourier transform values in the diffractive case are now along an arc when a two-dimensional representation is used. Experimentally, the wave scattered by a weakly scattering object can be recorded on the surface of an image sensor. Under the first-order Born approximation, this measured scattered field is interpreted as a two-dimensional spherical subset of the Fourier transform of the sample scattering potential distribution.

\begin{figure}[!t]
\centering
\includegraphics[width=0.48\textwidth]{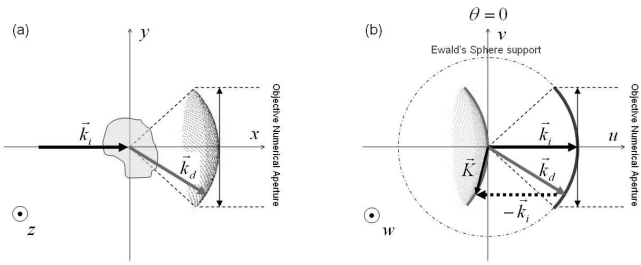}
\caption{Mapping representation of the object frequencies in the case of scattered light collected in transmission. (a) The object scattered frequencies collected through the objective numerical aperture for an incident plane wave of wave vector $\vec{k_i}$ lie on the surface of a spherical cap. (b) The Fourier transform of the object scattering potential are distributed over the surface of a spherical cap, the top of which coincides with the frequency origin.}
\label{fig:5}
\end{figure}

\begin{figure}[!t]
\centering
\includegraphics[width=0.48\textwidth]{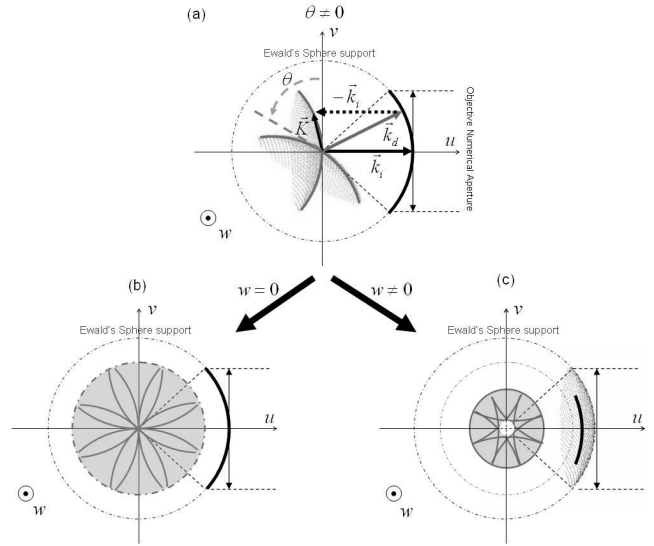}
\caption{Representation of the object frequency mapping obtained by object rotation along the $z$-axis. (a) Spherical caps of the object scattering frequencies mapped for $\theta = 0$ and $\theta\neq 0$. The extended support of the mapped frequencies is obtained by integrating over a large number of rotation angles. Slices of the extended supports are shown at (b) $w = 0$ and (c) $w\neq 0$. The regions of the frequency domain covered by rotating the object are shown in light grey; note the formation of a missing disk for $w\neq 0$, the surface of which increases with the distance from the frequency origin. These missing frequencies form a volume that resembles that of an apple core.}
\label{fig:6}
\end{figure}

This spherical subset of the Fourier transform is limited by the $NA$ of the experimental setup as illustrated in Figure \ref{fig:5}. By rotating the sample with a fixed illumination (constant vector $\vec{k_i}$), different subsets of the three-dimensional frequency distribution of the object scattering potential can be mapped (see Figure \ref{fig:6}(a)). In contrast to the non-diffractive case, the extended support of frequencies mapped by recording spatial frequencies for a large number of rotation angles is not a complete sphere. Indeed by rotating a sphere cap along an axis of rotation tangent to the cap top as shown in Figure \ref{fig:6}(a), one can cover a volume of points within a sphere except for the points near the rotation axis as depicted in Figure \ref{fig:6}(b,c). The bend of the cap does not allow for the mapping of the frequencies along the axis of rotation. Furthermore, when slices perpendicular to the rotation axis are examined, one find out that disks centered on the rotation axis are not mapped. The longer the distance from the frequency origin, the larger the disk area becomes that is not covered by the rotating caps. Accordingly, there exists a region of the frequency domain that is not mapped by the rotation. This missing volume in the frequency domain was named the \textquotedblleft missing apple core\textquotedblright\,after its characteristic shape represented in Figure \ref{fig:7}.

\section{Simulations}
\begin{figure}[!t]
\centering
\includegraphics[width=0.48\textwidth]{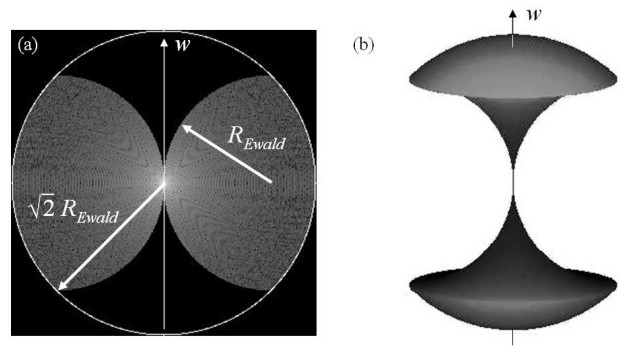}\caption{Representation of the missing apple core in the frequency domain as obtained in the case of an object rotation along the $z$-axis and measurements of forward scattered fields with $NA=1$. (a) Slice crossing the $w$-axis of the three-dimensional object frequencies. The missing frequencies are shown in black and the cut-off frequency of the low-pass filter is indicated by a white circle. (b) Three-dimensional representation of the missing frequencies forming a volume similar in shape with that of an apple core. The $w$-axis is the symmetry axis of the apple core and the $z$-axis is the axis of rotation of the object as shown in Figure \ref{fig:5}.}
\label{fig:7}
\end{figure}

In this section, the effects of the missing apple core on the reconstruction of ideal objects such as spheres, cylinders and membranes from the scattered fields recorded by sample rotation diffractive tomography are examined. From the shape of the apple core represented in Figure \ref{fig:7} where high frequencies are missing along the rotation axis, it can be inferred that the effects of the missing apple core should translate into blurring around image borders that are perpendicular to the axis of rotation.

A quantitative investigation into the effects of the missing apple core was conducted in the following way. Perfect three-dimensional objects were modeled and their three-dimensional Fourier transforms computed using Matlab standard functions. Then, the effect of the limited $NA$ of the microtomograhy setup was simulated. This was done by applying a spherical low-pass filter on the computed object frequencies that filters out the high frequencies not collected by the microscope objective (only low frequencies within a sphere are used). Next, the missing apple core region of the frequency domain was removed from the computed object frequencies. Finally the three-dimensional inverse Fourier transform of the remaining object frequencies after low-pass filtering and missing apple core removal was computed.

The mapping in the frequency domain of the object frequencies requires the knowledge of the radius of the Ewald sphere and the effective numerical aperture of the instrument which depend on the detailed of the experimental setup arrangement. Here, we assumed that light collection by the instrument is limited by $NA$ only as shown in Figure \ref{fig:5} and we chose $NA=1$ for convenience. Doing so, we compute an extreme case where the relative importance of the missing apple core is maximal. Indeed, for lower numerical aperture, the relative volume of the missing apple core to the volume of effectively recorded frequencies diminishes, thus reducing the apple core effect on the specimen reconstruction quality.

An $NA$ of 1 (in air) means that for each direction of illumination, the spherical cap collected on the image sensor is half of the corresponding $\sigma$ Ewald sphere (in our case, the spherical caps of the $\sigma$ spheres of Figure \ref{fig:3} that verify $\vec{k_i}\cdot\vec{k_d}\geq 0$). Under this circumstance, it is clear from Figure \ref{fig:3} that the cut-off frequency of the low pass filter is $\sqrt{2}R_{Ewald}$. Consequently the maximum of the simulated object frequencies should be set to be at least $\sqrt{2}R_{Ewald}$ so as to map all frequencies collected by the simulated instrument.

In the following, simulations have been conducted using a matrix with size of
$301\times 301\times 301$ pixels (a compromise between computation time and reconstruction quality) and our cut-off frequency of $\sqrt{2}R_{Ewald}$ was set to 150 pixels as shown in Figure \ref{fig:7}. The matrix values used to define the object and the background were set to 1 and 0, respectively. For all simulations, the $z$-axis was the rotation axis of the object and the $w$-axis the axis of revolution of the apple core in the frequency domain.

The effect of the missing apple core was examined on three different kinds of object shapes, namely solid spheres, solid cylinders and hollow spheres depicting membrane-like objects like vesicles or cell membranes. The geometry of the sphere provides border perpendicular to each possible direction in the object space, so that the comparison between the effect of the three-dimensional low-pass filter and the effect of the missing apple core is made possible for all directions. Two cylinders with their axes of revolution perpendicular and identical with the axis of rotation were selected to observe the effect of the missing apple core on large surfaces perpendicular to the axis of rotation. Finally an empty sphere mimicking membrane objects such as cells was simulated.\begin{figure}[!t]
\centering
\includegraphics[width=0.48\textwidth]{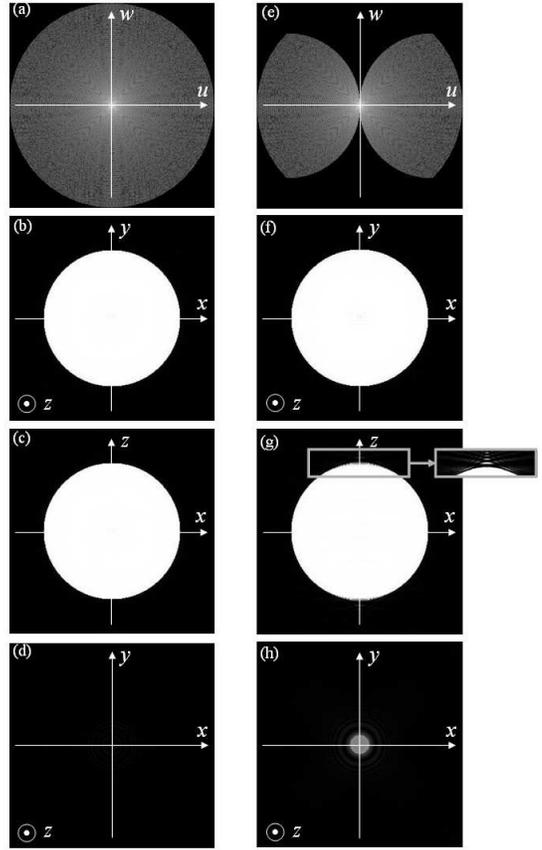}
\caption{Simulated effects of the missing apple core on the reconstruction of a sphere. (a) Slice in the $uw$-plane of the complete object frequencies, that is, without the missing apple core but after applying the low pass filter. Slices in the (b) $xy$-plane and (c) $xz$-plane of the reconstructed sphere obtained by inverse Fourier transform of the sphere frequencies as shown in figure (a). (d) Slice of the reconstructed sphere taken at constant $z$ at one of the extremities of the sphere. (e) Slice in the $uw$-plane of the incomplete object frequencies, that is, with the missing apple core and after applying the low pass filter. Slice in the (f) $xy$-plane and (g) $xz$-plane of the reconstructed sphere obtained by inverse Fourier transform of the sphere frequencies as shown in figure (e). The inset of figure (g) shows the sphere extremity using an enhanced contrast. Note the deformations in the reconstructed sphere from the incomplete frequency information. Particularly noticeable are deformations in the $z$-direction corresponding to the rotation axis of the object which creates the missing apple core along the $w$-axis in the frequency domain. (h) Slice taken at constant $z$ corresponding to the extremity of the sphere shows a ring pattern generated by the missing spatial frequencies of the apple core. The object edges perpendicular to the $z$-axis are the most blurred.}
\label{fig:8}
\end{figure}

Simulation results for a solid sphere are reported in Figure \ref{fig:8}. The original sphere had a radius of $100$ pixels and the sphere frequencies after applying the low-pass filter are shown in Figure \ref{fig:8}(a). The reconstruction of the solid sphere from the frequencies of Figure \ref{fig:8}(a) is shown in Figure \ref{fig:8}(b-d). This reconstruction represents the best possible result for microtomography limited by the $NA$ only. Figure \ref{fig:8}(e) shows the frequencies of the same object after removing the apple core region in the frequency domain and Figure \ref{fig:8}(f-h) the corresponding reconstruction by inverse Fourier transform. In contrast to the isotropic blurring effect of the low-pass filter, the effect of the missing apple core was chiefly observed on the two sphere tops perpendicular to the axis of rotation as evidenced by the comparison of Figure \ref{fig:8}(c) and \ref{fig:8}(g).

In addition to the border blurring, artefacts appeared in the background near the sphere tops perpendicular to the axis of rotation. A close look in Figure \ref{fig:8}(d,h) at the region near a sphere top perpendicular to the rotation axis revealed that if the low-pass filter changed the background level of the reconstructed object, the missing apple core generated grey disks that did not belong to the original sphere.

\begin{figure}[!t]
\centering
\includegraphics[width=0.48\textwidth]{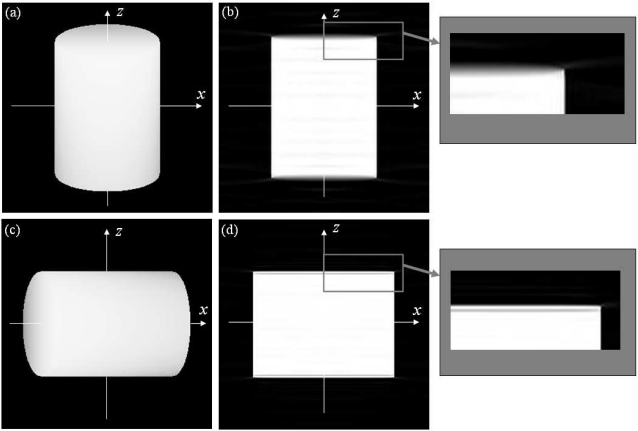}
\caption{Simulated effects of the missing apple core on the reconstruction of cylinders with main axis (a,b) parallel and (c,d) perpendicular to the $z$-axis of rotation. The object frequencies comprised in the apple core of figure \ref{fig:7} were removed and then the objects were computed by inverse Fourier transform. Three-dimensional representations of the original objects are shown in figure (a) and (c). Slices in the $xz$-plane of the reconstructed cylinders from their object frequencies with the missing apple core are shown in figure (b) and (d). Note the deformation along the $z$-axis in figure (b) and (d).}
\label{fig:9}
\end{figure}

Simulation results for the solid cylinders are shown in Figure \ref{fig:9}. The two cylinders had a radius of $75$ pixels and a height of $100$ pixels. The axis of symmetry of the first cylinder shown in Figure \ref{fig:9}(a,b) coincides with the rotation axis and the axis of the second cylinder shown in Figure \ref{fig:9}(c,d) is perpendicular to the axis of rotation. The original cylinders are shown on the left side of Figure \ref{fig:9} and the reconstructed cylinders on the right side. Blurring of the cylinder borders perpendicular to the axis of rotation is observed in the zooms of Figure \ref{fig:9}(c,d). The extent of border blurring is larger when the cylinder symmetry axis coincides with the rotation axis because in this case the surface border perpendicular to the rotation axis possesses high frequencies along the rotation axis that are filtered out due to the missing apple core.

\begin{figure}[!t]
\centering
\includegraphics[width=0.48\textwidth]{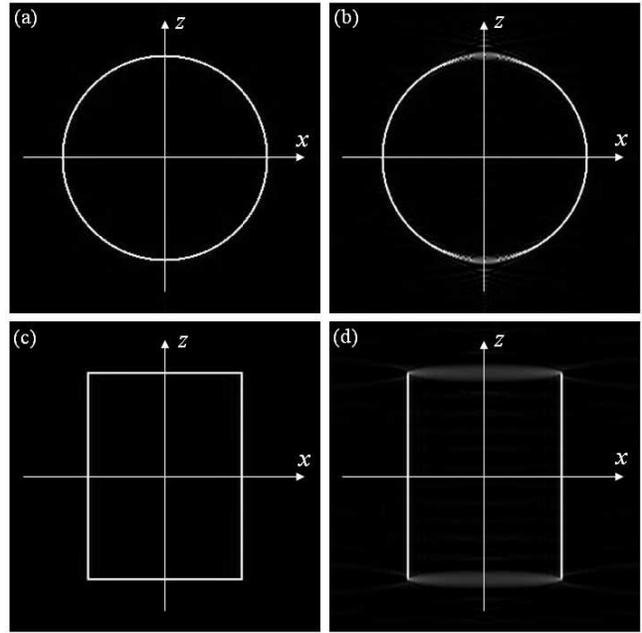}
\caption{Simulated effects of the missing apple core on the reconstruction of membrane-type objects having (a,b) a spherical shape and (c,d) a cylindrical shape. The inner and outer of the membranes have the same properties. The figures (a,c) show slices crossing the symmetry axis of the original objects and the figures (c,d) show the same slices for the reconstructed objects. The reconstruction is done by inverse Fourier of the original object frequencies with a missing apple core and after application of a low pass filter.}
\label{fig:10}
\end{figure}

Simulation results for membranes that have a similar structure with biological samples are shown in Figure \ref{fig:10}. Simulations were performed on two types of membranes, namely spherical and cylindrical. The spherical membrane had a membrane thickness of two pixels defined by the $99$ and $100$ pixels. The cylindrical membrane consisted of two cylinders forming a membrane with thickness of $2$ pixels. The reconstructed membranes of Figure \ref{fig:10}(c,d) show that the blurring of the borders perpendicular to the rotation axis is more pronounced that in the case of the solid objects of Figures \ref{fig:8} and \ref{fig:9}. Particularly the surfaces perpendicular to the axis of rotation of the cylinder membrane are poorly defined. Moreover, the reconstructed backgrounds exhibit variations that did not belong to the original objects.

\begin{figure}[!t]
\centering
\includegraphics[width=0.48\textwidth]{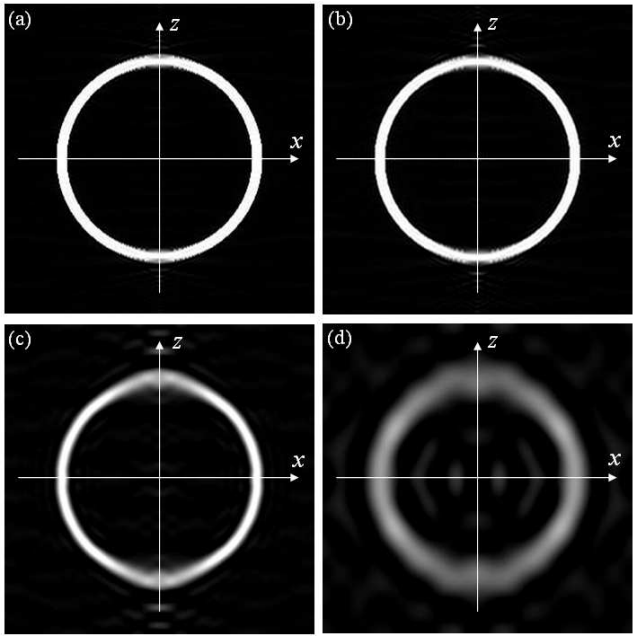}
\caption{Simulated effects of the elongation due to the missing apple core on the reconstruction of a membrane-type object having a spherical shape with a membrane thickness corresponding to $10$ pixels. The inner and outer of the membrane have the same properties. Different membrane sizes of $44$, $31.5$, $9.5$ and $3\,\mu m$ were used for figures (a), (b), (c) and (d), respectively. The reconstruction is obtained by inverse Fourier of the original object frequencies with a missing apple core and after application of a low pass filter. The figures (a,b,c,d) show slices of the reconstructed object in a plane containing the rotation axis $z$. The reconstructed membrane object shows an elongation along the $z$-axis, the effect being pronounced for small objects.}
\label{fig:11}
\end{figure}

Finally, the effect of the object size on the reconstruction of spherical membrane-like objects is investigated by fixing the illumination light wavelength at $633\,nm$ and representing different object sizes by the same number of pixels. The physical radius of the Ewald sphere being defined by the wavelength, it is therefore kept constant, but numerically represented by a different number of pixels in the frequency domain. Figure \ref{fig:11} shows the simulation results for the object sizes of $44$, $31.5$, $9.5$ and $3\,\mu m$, corresponding to the Ewald sphere radii of $105$, $71$, $21$ and $7$ pixels. For large sizes (Fig. \ref{fig:11}(a-b)), the relative influence of the missing frequencies decreases resulting in very good reconstructions. For small sizes (Fig. \ref{fig:11}(c-d)), however, an elongation of the reconstructed object along the $z$-axis can be observed. Note that this elongation becomes more and more apparent as the object size is decreased. This effect is very similar to the effect of the missing-cone in transmission microscopy or fluorescence microscopy, which influence becomes more and more visible for smaller object, and which precludes an efficient specimen reconstruction, even after deconvolution \cite{Chomik1997}. The main difference is that in tomography with specimen rotation, the axis of lower quality image reconstruction corresponds to the physical rotation axis of the specimen, while in conventional microscopy, the optical axis is the direction of worst imaging conditions.

Note however that the observed elongation along the rotation axis in diffractive tomography with sample rotation is less pronounced than that observed in standard transmission microscopy or in fluorescence microscopy along the optical axis \cite{Chomik1997,Colicchio2005}. This can be easily understood if one considers the so-called missing-cone in transmission or fluorescence microscopy \cite{Streibl1985}: while high frequencies are captured along lateral axes, no frequencies are measured along the optical axis, and even the maximum thickness of the captured frequency support along this direction is several times smaller than its lateral extension. Note that this frequency support has the same shape in diffractive tomography with variable illumination \cite{Wolf1969,Dandliker1970,Lauer2002}. On the contrary, in diffractive tomography with sample rotation, the relative volume of the missing apple core compared to the captured frequency support is rather small (see Fig. \ref{fig:7}), which explains the good quality of the simulated reconstructions, even if, strictly speaking, the obtained images do not exhibit isotropic properties.

\section{Conclusion}
The mapping of the observed object frequencies in diffraction tomography using coherent light with sample rotation along one axis does not allow for a complete mapping of the frequency along the axis of rotation due to the finite Ewald sphere radius. Under the first-order Born approximation, the projection of the diffracted waves in the frequency space onto the set of sphere caps covered by the sample rotation featured a missing part that had a characteristic shape resembling that of an apple core. The main effects of this missing apple core, as revealed by numerical reconstructions of ideal objects, is the presence of a more important blur of the reconstructed object borders perpendicular to the axis of rotation, a slight modification of the background close to these borders, and a slight elongation along the rotation axis. The later effect is similar to the elongation observed in transmission microscopy along the optical axis, but less noticeable. This elongation effect is found more pronounced for objects having sizes of the order of the wavelength of the illumination light. We also noticed that the effects were more pronounced for membrane-type objects than for solid objects.

\end{document}